\documentclass[10pt]{article}

\usepackage[utf8]{inputenc}
\usepackage[T1]{fontenc}
\usepackage{lmodern}
\usepackage[margin=1in]{geometry}
\usepackage{graphicx}
\usepackage{xcolor}
\usepackage{enumitem}
\usepackage{booktabs}
\usepackage{listings}
\usepackage{hyperref}
\usepackage[round,authoryear]{natbib}
\usepackage{tikz}
\usepackage{pgfplots}
\pgfplotsset{compat=1.18}
\usetikzlibrary{shapes,arrows.meta,positioning,calc}
\usepackage{subcaption}

\hypersetup{
  colorlinks=true,
  linkcolor=blue,
  citecolor=blue,
  urlcolor=blue
}

\lstdefinelanguage{yaml}{
  keywords={true,false,null,y,n},
  keywordstyle=\color{blue}\bfseries,
  basicstyle=\ttfamily\small,
  sensitive=false,
  comment=[l]{\#},
  morecomment=[s]{/*}{*/},
  commentstyle=\color{gray}\itshape,
  stringstyle=\color{teal},
  morestring=[b]',
  morestring=[b]"
}

\title{Jas: AI-Paired Engineering as a Revival of N-Version Programming}

\author{Jason Hickey\\
Independent\\
\texttt{jasonh@gmail.com} \textperiodcentered{} \url{https://github.com/jyh/jas}\thanks{Work performed on personal time, independent of the author's employer. Views and opinions are the author's own and do not represent any organization.}}

\date{}

\begin{document}

\maketitle

\begin{abstract}
I report a case study in AI-paired software engineering: five working ports
of a vector illustration application across Rust, Swift, OCaml, Python, and
browser-based platforms, built by a single developer in approximately 120
evening hours. The methodology pairs AI-assisted implementation with two
safeguards --- a precise executable YAML specification serving as the single
source of truth, and parallel implementations functioning as a built-in
differential-testing layer. The five ports share a 23{,}000-line
specification; per-port native code ranges from 0 to roughly 95{,}000 lines,
reflecting the specification's escape hatch. I argue that AI-paired
engineering, conditional on these two safeguards, makes feasible scope of
work that conventionally requires multiple developer-years, and frame the
methodology as a revival of N-version programming, a 1980s approach
abandoned on cost grounds that AI changes. The paper reports concrete
artifacts and honest limitations of the single-developer case study.
\end{abstract}

\section{Introduction}

Mature vector illustration applications --- exemplified by Adobe Illustrator
\citep{pfiffner2003adobe} and Inkscape \citep{inkscape} --- represent decades of
team development. I have used vector illustration applications, primarily
Adobe Illustrator, since 1990, for engineering drawings in my professional
work and for visual art outside it. Such applications have long seemed to
put extending them --- adding features, porting to new platforms --- out of
reach for individuals. This project began as a test of whether AI-paired
engineering could change that.

The artifact is five working implementations of a vector illustration
application, sharing a single executable specification and developed by one
developer over approximately 120 evening hours across seven calendar weeks.
The five implementations span Rust/Dioxus, Swift, OCaml, Python/PySide6,
and a browser-based Flask sketch --- five languages, five paradigms, five UI
frameworks. The shared specification is approximately 23{,}000 lines of
YAML; per-port native code ranges from a few thousand to roughly 95{,}000
lines, reflecting a specification escape hatch for platform-specific
concerns.

This paper reports the methodology. Its central claim is that AI-paired
engineering, when paired with two specific safeguards, makes feasible for a
single developer a body of work that would conventionally require multiple
developer-years. The safeguards are:

\begin{enumerate}[leftmargin=*]
\item \textbf{A precise, executable specification} serving as the single
  source of truth. The specification consolidates design decisions into
  one artifact and is rendered into all $N$ implementations by shared and
  per-port renderer code. Its cost is paid once and amortizes sub-linearly
  across $N$ implementations.
\item \textbf{Parallel implementations functioning as a built-in correctness
  check.} Each port stress-tests the others and exposes places where the
  specification is underspecified. Per-port cost is linear in $N$;
  correctness gain is sub-linear but real --- particularly in catching the
  visual and behavioral divergences that automated tests miss.
\end{enumerate}

I argue these two safeguards can be understood as a revival of N-version
programming \citep{avizienis1985nversion} --- a 1980s methodology that
called for multiple independent implementations to improve reliability, but
was largely abandoned because the cost of producing $N$ independent
implementations exceeded the reliability benefit. AI fundamentally changes
that economic argument. With AI handling most of the per-port mechanical
work, $N$ implementations of a single specification become feasible for a
single developer, and the resulting differential-testing layer makes the
productivity claim defensible without sacrificing correctness.

\paragraph{Contributions.} This paper contributes:
\begin{itemize}[leftmargin=*]
\item A case study of a single developer producing five platform-spanning
  implementations of a complex desktop application across seven weeks of
  evening work.
\item A specific methodology pairing an executable specification with
  parallel implementations as a differential-testing layer.
\item A field report on AI-paired software engineering in practice,
  including concrete prompts, persistent memory patterns, delegation
  strategies, and honest failure modes.
\item An open-source artifact (\url{https://github.com/jyh/jas}) and a
  reusable manual-testing protocol.
\end{itemize}

\paragraph{Paper structure.} Section~\ref{sec:setup} describes the project
setup. Section~\ref{sec:spec} introduces the executable specification
(Condition 1). Section~\ref{sec:nimpl} reports on parallel implementations
as a correctness check (Condition 2). Section~\ref{sec:ai} is a field
report on AI-paired engineering in practice. Section~\ref{sec:evidence}
presents the evidence supporting the scope claim. Section~\ref{sec:limits}
discusses limitations, Section~\ref{sec:related} positions the work in the
literature, and Section~\ref{sec:conclusion} concludes.

\section{Setup}
\label{sec:setup}

The project began with a deliberate choice to span five platforms rather
than focus on one. Two motivations drove this. First, multi-platform
implementation served as a test of the methodology: if the same
specification could drive five visibly equivalent ports across five
disparate UI frameworks, the specification was load-bearing enough to be
useful, not just a documentation artifact. Second, multiple implementations
provided what would become this paper's second condition --- a built-in
correctness check. The five platforms were selected to span paradigms,
ownership models, and runtime characteristics.

\paragraph{The five implementations.}
\begin{itemize}[leftmargin=*]
\item \emph{jas\_dioxus} (Rust, Dioxus framework) offers strict memory and
  concurrency guarantees and targets a high-performance web application.
  Rust's policy strictness was hypothesized to be a stress test for AI code
  generation; in practice it became the largest port by line count but not
  the most difficult to develop.
\item \emph{JasSwift} (Swift, SwiftUI) targets macOS and iOS, providing
  hardware-accelerated rendering and platform-native UI conventions.
\item \emph{jas\_ocaml} (OCaml) prioritizes safety and explicit interface
  management. The author has long experience in
  OCaml\footnote{The author co-authored \emph{Real World OCaml}, O'Reilly
  Media.} and treated this implementation as a control case for whether
  AI-paired engineering could match the speed of expert-language
  development.
\item \emph{jas} (Python, PySide6) prioritizes development speed and serves
  as a desktop reference using Qt's mature cross-platform widget set.
\item \emph{jas\_flask} (Python + Flask + HTML/JavaScript) is a
  server-rendered reference and early UI sketching ground. Its behavior is
  fully YAML-driven, with no per-feature server code, and it is not
  feature-complete in the sense the four native ports are.
\end{itemize}

\paragraph{App scope.} The application implements a substantial subset of
the feature set found in mature vector illustration applications: 27 tools
(Pen, Pencil, Paintbrush, Blob Brush, Eyedropper, Magic Wand, Lasso, Path
Eraser, Hand, Zoom, Selection, transforms such as Scale and Rotate, and
shape tools), 14 panels (Color, Swatches, Layers, Stroke, Brushes,
Character, Paragraph, Align, Artboards, Opacity, Gradient, Boolean,
Properties, and Magic Wand), and 22 dialogs (Color Picker, Document Setup,
Print Preferences, Brush Options, Hyphenation, Justification, and others).
The application supports vector paths, text with paragraph and character
styling, layers, transforms, undo, document save and restore, and PDF
export. Across the five ports the project includes approximately 4{,}600
automated test functions and 36 manual-test transcript files; the Color
Panel alone defines 98 numbered manual scenarios.

\paragraph{Comparison anchors.} Adobe Illustrator has been developed
continuously since 1987, representing roughly 39 years of large-team work.
Inkscape, the leading open-source comparison, has been under continuous
development since 2003. Commercial alternatives such as Affinity Designer
and Figma also represent multi-developer-year efforts by sustained teams.
The artifact described in this paper does not match these in feature
completeness --- it lacks gradient mesh, advanced text shaping at the level
of professional typesetting engines, raster effects, full SVG round-trip
fidelity, plugins, and color management beyond basic profiles. Its
contribution is not in matching mature applications, but in demonstrating
that a substantial subset is achievable by a single developer in
dramatically less time under a specific methodology.

\section{The executable specification}
\label{sec:spec}

\subsection{What the specification is}

The specification is approximately 23{,}000 lines of YAML organized as a
directory tree under \texttt{workspace/}. It declares the application's
panels, dialogs, tools, menus, keyboard shortcuts, theme tokens, and
document state model.\footnote{The full source for the specification and
all five implementations is available at \url{https://github.com/jyh/jas}.}
It is \emph{executable} in the sense that each implementation contains a
generic interpreter that reads the YAML at startup and constructs working
UI directly from it, rather than treating the YAML as documentation that
has been re-encoded in each port.

Each declarative construct in the YAML has a corresponding native renderer
in each port. A \texttt{container} becomes a \texttt{VBox} in PySide6, a
\texttt{VStack} in SwiftUI, a \texttt{<div>} in HTML, a Cairo layout group
in OCaml's GTK binding, and a \texttt{dioxus::div} in Rust. A
\texttt{number\_input} becomes the platform's native numeric-input widget.
Behavioral semantics --- bidirectional bindings, dialog state with get/set
lambdas, action dispatch, slider snap-on-write, theme-aware styling --- are
also part of the YAML and are evaluated by the same interpreter that
constructs the widgets.

This approach has a lineage in executable specifications for language
semantics, discussed in Section~\ref{sec:related}. The contribution here is
to apply the pattern to interactive application UI: not just the static
structure of the interface, but its reactive behavior under user input.

\subsection{The shared interpreter and the escape hatch}

Each port carries two layers of code: a shared interpreter that loads YAML
and dispatches it through generic renderers, and a per-port escape hatch
for platform-specific concerns that the generic renderers cannot
adequately express. The shared interpreter is approximately 12{,}500 lines
(originating in Python and reused or ported to the other languages); the
per-port renderer layers vary substantially in size, reflecting how often
each platform requires native code that the generic dispatch cannot
produce. The interpreter is itself an artifact of the project --- a
reusable, language-agnostic engine for evaluating a declarative UI
specification --- and represents a non-trivial portion of the project's
design work.

The escape hatch is not a flaw in the methodology; it is the methodology's
load-bearing flexibility. Custom canvas widgets, hardware-accelerated
drawing surfaces, platform-specific gesture handling, and complex state
synchronization with native UI frameworks (such as SwiftUI's reactive
\texttt{@ObservedObject} model) all fall outside what YAML can practically
describe. The discipline is that everything \emph{that can} be expressed
in YAML is expressed in YAML; native code exists only where the
specification's expressive power runs out.

\subsection{Running example: the Color Panel}
\label{sec:colorpanel}

The Color Panel illustrates the architecture in a compact form. The YAML
specification for the Color Panel comprises four files:
\begin{itemize}[leftmargin=*]
\item \texttt{workspace/panels/color.yaml} (493 lines) --- the panel
  layout, slider rows by mode (HSB / RGB / CMYK / Grayscale / Web Safe),
  fill-stroke widget binding, recent-colors strip, mode buttons, hamburger
  menu
\item \texttt{workspace/dialogs/color\_picker.yaml} (250 lines) --- the
  modal color picker with hex field, 2D gradient, hue bar, channel inputs,
  color-swatches link
\item \texttt{workspace/templates/color\_picker\_fields.yaml} (37 lines)
  --- reusable HSB / RGB / CMYK row template
\item \texttt{workspace/templates/fill\_stroke\_widget.yaml} (110 lines)
  --- the small fill / stroke selector widget
\end{itemize}

Total: 890 lines of declarative YAML, written once and consumed by all
five implementations.

The per-port native code dedicated to the Color Panel ranges from zero to
over a thousand lines. The OCaml port carries no dedicated color-panel
code: the generic YAML interpreter is sufficient. The Swift port adds 59
lines of state-bridging code (\texttt{ColorPanelSync.swift}) to mediate
between the YAML-driven state model and SwiftUI's reactive update cycle.
The Python port adds approximately 123 lines for a custom color-bar widget
painted with QPainter. The Rust port adds approximately 1{,}300 lines
spread across three files (\texttt{color\_panel\_view.rs},
\texttt{color\_panel.rs}, \texttt{fill\_stroke\_widget.rs}) implementing
the gradient widget, the hue bar, and a custom fill-stroke composite ---
immediate-mode rendering that does not fit the Dioxus declarative idiom.

This distribution is informative. The platforms whose UI idioms are
closely aligned with the YAML's declarative model require very little
native code; the platforms where the model collides with a different
rendering paradigm require more. The amount of native code per port is, in
effect, a measure of how well the specification's expressive power matches
the target framework. Section~\ref{sec:nimpl} returns to this distribution
as a correctness benefit, as cross-port comparison reveals where the
specification is underspecified.

\begin{figure}[h]
\centering
\begin{lstlisting}[language=yaml]
content:
  type: container
  id: cp_content
  layout: column
  style: { padding: 4, gap: 6 }
  children:

    # Row 1: Fixed swatches | rule | Recent colors
    - type: container
      id: cp_swatches_row
      layout: row
      style: { gap: 2, alignment: center }
      children:

        - id: cp_none_swatch
          type: icon_button
          icon: color_none
          summary: "None"
          style: { size: 16 }
          behavior:
            - event: click
              action: set_active_color_none

        - id: cp_black_swatch
          type: color_swatch
          summary: "Black"
          style: { size: 16 }
          bind:
            color: "#000000"
          behavior:
            - event: click
              action: set_active_color
              params: { color: "#000000" }
\end{lstlisting}
\caption{An excerpt from \texttt{workspace/panels/color.yaml} showing the
top-level container, two concrete widgets (an icon button and a color
swatch), their bindings, and click behaviors. The full file is 493 lines;
this excerpt illustrates the declarative style.}
\label{fig:yaml}
\end{figure}

\subsection{Sub-linear cost across $N$ implementations}

The headline implication: five working Color Panel implementations from
890 lines of shared YAML, with native code totaling roughly 1{,}500 lines
across four ports. The fifth is fully YAML-driven.

In conventional cross-platform development, each port carries the full
conceptual load of a feature independently. Color picker logic, slider
snapping, mode-button state, bidirectional channel bindings --- all of
these would be reimplemented in each language and framework. The
specification consolidates these decisions into a single artifact. When a
new feature is added (a new slider mode, a new dialog field, a new
keyboard shortcut), it is added to the YAML once, and propagates to all
five ports through their interpreters. When a behavioral detail is refined
--- for example, the HSB-degenerate-at-S=0 case discussed as a vignette in
Section~\ref{sec:nimpl} --- it is refined in one place.

The cost of the specification is paid once and amortizes sub-linearly
across $N$ implementations. The cost of per-port renderer code is paid per
port, but is much smaller than the cost of a full per-port implementation
because the renderer code only implements the escape hatch --- the part
the spec cannot express. Across the project, native code totals
approximately 300{,}000 lines spread across five ports; the shared
specification plus interpreter totals approximately 35{,}000 lines. The
interpretation is not that the project is ``8.5 times smaller'' --- much
of the native code is platform glue with no YAML counterpart --- but that
the specification carries the conceptual work, and the per-port code
carries the platform-specific machinery.

\begin{figure}[h]
\centering
\begin{tikzpicture}
\begin{axis}[
  xbar stacked,
  width=0.85\textwidth,
  height=5.5cm,
  bar width=0.45cm,
  xlabel={Lines of code},
  symbolic y coords={Flask, Rust, Python, Swift, OCaml},
  ytick=data,
  enlarge y limits=0.18,
  xmin=0,
  xmax=2400,
  legend style={at={(0.5,-0.28)}, anchor=north, legend columns=-1,
                font=\small, /tikz/every even column/.append style={column sep=0.5cm}},
  tick label style={font=\small},
  label style={font=\small},
  nodes near coords align={horizontal}
]
\addplot[fill=blue!30, draw=blue!50!black]
  coordinates {(890,Flask) (890,Rust) (890,Python) (890,Swift) (890,OCaml)};
\addplot[fill=orange!60, draw=orange!50!black]
  coordinates {(0,Flask) (1309,Rust) (123,Python) (59,Swift) (0,OCaml)};
\legend{Shared YAML (890), Per-port native code}
\end{axis}
\end{tikzpicture}
\caption{Color Panel spec amortization. Shared YAML (890 lines) drives
five working implementations; per-port native code varies from 0 (OCaml,
fully YAML-driven; Flask, fully server-side YAML) to 1{,}309 (Rust, with
custom canvas widgets for the gradient, hue bar, and fill-stroke
composite). Specific values: OCaml 0, Swift 59, Python 123, Rust 1{,}309,
Flask $\sim$0.}
\label{fig:amortization}
\end{figure}
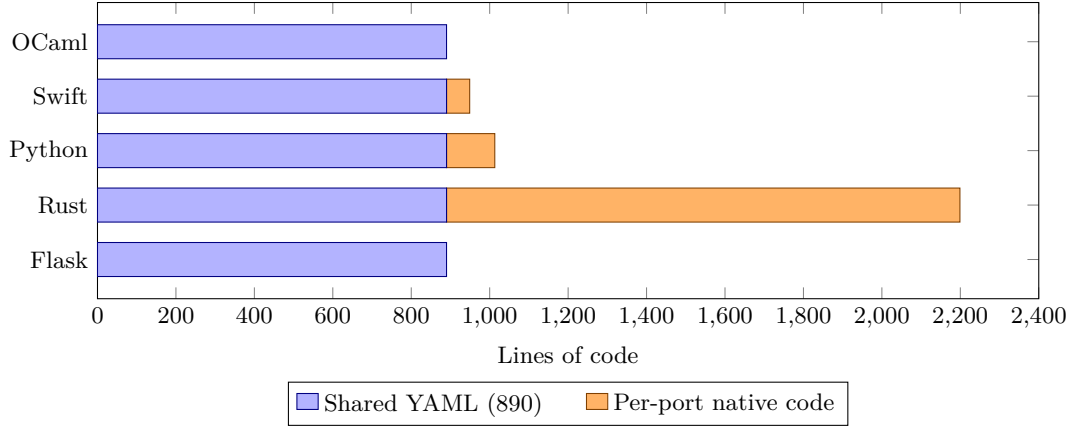

\section{Parallel implementations as correctness check}
\label{sec:nimpl}

\subsection{$N$ implementations as differential testing}

The five implementations exist not only to demonstrate that the
specification is portable, but to make divergence between them observable.
Two ports producing visibly different output from the same specification
reveal a flaw --- either in the specification (it permits more than one
reasonable interpretation), in one of the implementations (it has a bug),
or in both. The shared specification provides the contract; the parallel
implementations provide the cross-check.

This is differential testing \citep{mckeeman1998differential} applied to
interactive application code. Differential testing has succeeded most
prominently in compiler validation \citep{yang2011csmith}, where multiple
independent compilers already exist as targets to compare against. In this
work, the implementations are not independent --- they are derived from a
shared specification --- but they are independent enough in their use of
native platform features (rendering, layout, event handling, state
management) that bugs in either the spec or the implementations surface
when output diverges.

In practice, much of this differential testing was performed manually
rather than automated. For each user-visible feature, a transcript file
(e.g., \texttt{transcripts/COLOR\_TESTS.md}) lists numbered scenarios that
an operator runs in each port and marks with a pass date. Automated tests
catch state-level divergence; manual transcripts catch visual, timing, and
behavioral divergence that automated tests miss.

\subsection{Spec underspecification: bugs found by divergence}
\label{sec:vignettes}

The most informative divergences are not implementation bugs but
specification gaps. The specification permits a reasonable interpretation,
two ports adopt different reasonable interpretations, and the divergence
forces the specification to be sharpened. Five examples from the Color
Panel illustrate the pattern.

\paragraph{Hue collapses to zero when saturation drops to zero.} When the
saturation slider is dragged to zero in HSB mode, the hue channel becomes
mathematically meaningless --- the color is gray at any hue. The Python
and Rust ports initially snapped hue to zero when saturation reached
zero, a literal reading of the HSB $\rightarrow$ RGB conversion. The
OCaml and Swift ports preserved the prior hue value. The user-visible
effect: dragging saturation to zero and back up in Python or Rust
produced a red, not the green the user had been working with. Cross-port
testing surfaced this within the first hour of manual Color Panel
testing. The specification was updated to require explicit preservation
of degenerate-channel values.

\paragraph{CMYK channels collapse when K reaches 100.} A parallel case in
the CMYK model. At K=100, the displayed color is black regardless of C,
M, and Y. Three of the four native ports initially zeroed C/M/Y when K
crossed 100. The fourth preserved them. The user-visible effect: dragging
K back down from 100 should restore the working color, not produce black.
The specification was updated to require channel preservation on
degenerate transitions.

\paragraph{Web Safe RGB snap timing.} The Web Safe RGB mode constrains
channel values to multiples of 51. Initial implementations differed on
\emph{when} the snap should occur: on slider drag, on slider release, on
widget write-back, or on display only. Two ports snapped on display and
produced incorrect committed values; one snapped on drag and produced
jumpy visual feedback. The specification was updated to require
snap-on-write-back: the value visible during drag is the snapped value,
and the committed value matches what is shown.

\paragraph{Hex commit reverts on dialog OK.} The modal color picker dialog
exposes a hex field. When the user typed a new hex value and pressed OK,
three of four ports correctly applied the new color, but one (Python)
reverted to the previous value. The cause was that the dialog's
evaluation context had captured a snapshot of the hex field before the
typed-in value flowed through the get/set lambda chain. The specification
required clarification on when the dialog evaluation context must be
rebuilt against typed-in values rather than the render-time snapshot.

\paragraph{Recent colors push during slider drag.} Recent colors should be
added on commit (slider release, hex Enter, swatch click), not during
drag. Two ports initially added every intermediate drag value to the
recent list, producing a stream of slight variations of the same color.
The specification was updated to make commit semantics explicit and to
require ports to push only on commit events.

These divergences are not anomalies; they are the methodology working as
designed. Each one would have been silently wrong in a single
implementation. With five implementations and manual cross-checking, each
was visible within hours of the relevant manual test session.

\begin{figure}[h]
\centering
\begin{subfigure}[t]{0.19\textwidth}
  \centering
  \includegraphics[width=\linewidth]{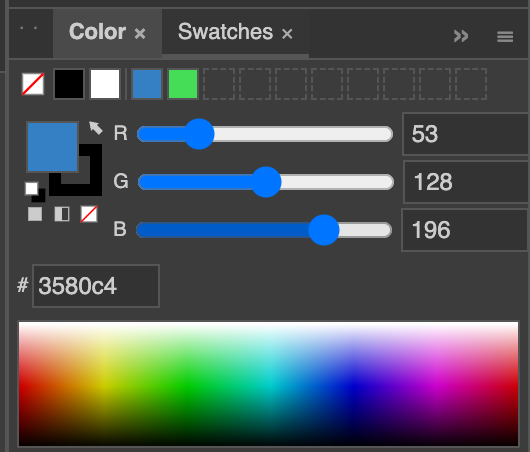}
  \caption{jas\_dioxus (Rust)}
  \label{fig:cp-rust}
\end{subfigure}\hfill
\begin{subfigure}[t]{0.19\textwidth}
  \centering
  \includegraphics[width=\linewidth]{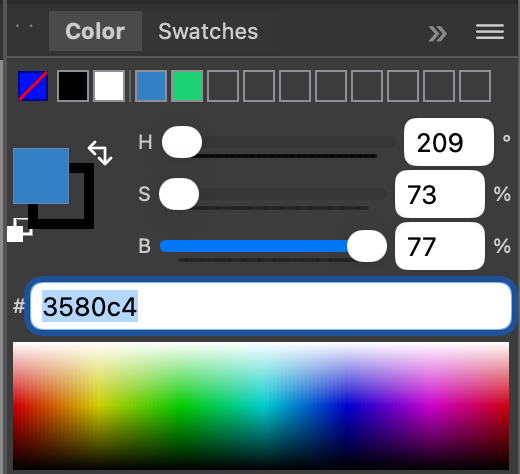}
  \caption{JasSwift}
  \label{fig:cp-swift}
\end{subfigure}\hfill
\begin{subfigure}[t]{0.19\textwidth}
  \centering
  \includegraphics[width=\linewidth]{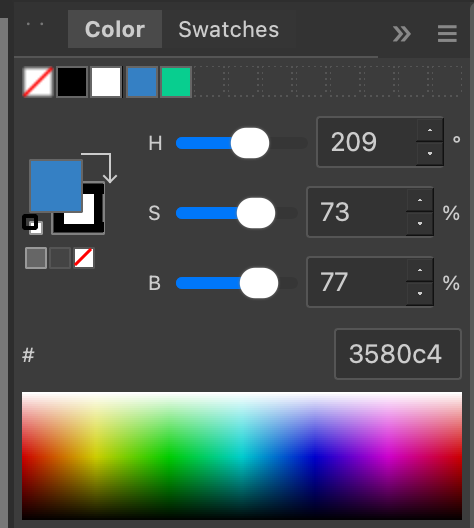}
  \caption{jas\_ocaml}
  \label{fig:cp-ocaml}
\end{subfigure}\hfill
\begin{subfigure}[t]{0.19\textwidth}
  \centering
  \includegraphics[width=\linewidth]{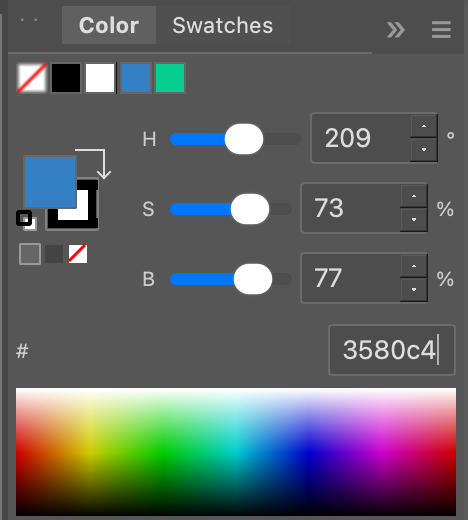}
  \caption{jas (Python)}
  \label{fig:cp-python}
\end{subfigure}\hfill
\begin{subfigure}[t]{0.19\textwidth}
  \centering
  \includegraphics[width=\linewidth]{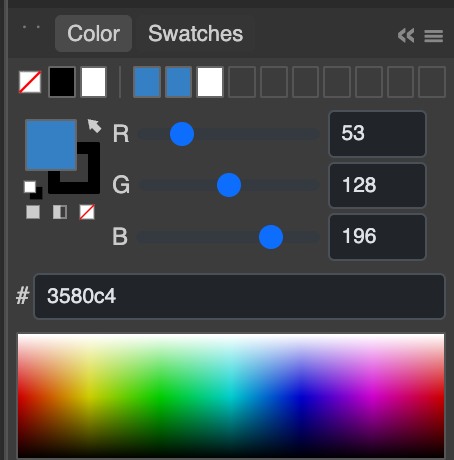}
  \caption{jas\_flask}
  \label{fig:cp-flask}
\end{subfigure}
\caption{The Color Panel rendered in all five ports, each showing the
active color \texttt{\#3580c4} on the Dark Gray theme. All five
implementations are driven by the same 890 lines of YAML specification.
Visible-by-inspection equivalence is the figure's intent: the active
color swatch, recent-color strip, slider arrangement, hex field, and
color bar appear in equivalent layout positions across the ports, with
platform-native variation in slider styling, spinbox controls, and tab
presentation. The bugs documented in Section~\ref{sec:vignettes} were
caught by precisely this kind of side-by-side comparison.}
\label{fig:screenshots}
\end{figure}

\subsection{Cost and correctness trade-offs}

The cost of $N$ implementations is linear in $N$: each port requires its
own renderer code, its own manual testing pass, its own propagation of bug
fixes. In this project, total manual testing across five ports accounts
for the largest share of remaining developer time, more than spec writing
or per-port implementation.

The correctness benefit is sub-linear. Most spec-completeness bugs surface
between $N=1$ and $N=2$ --- the moment a second implementation reveals
that the first made an unstated choice. $N=3$ catches a residual layer of
subtler bugs that depend on three-way comparison (one port agrees with
the spec on a subtle reading, two ports diverge in different directions).
Beyond $N=3$, additional implementations primarily verify that
earlier-caught spec refinements have been correctly propagated, and serve
as protection against regression on previously-found classes of bugs.

For this project, $N=5$ was chosen for reasons beyond differential testing
--- diversity of platforms, languages, and frameworks --- but a smaller
$N$ (3 or 4) would likely have provided most of the correctness benefit at
proportionally lower cost. The right $N$ is a project-specific decision
driven by platform coverage requirements as much as by differential
testing economics.

\subsection{Connection to N-version programming}
\label{sec:nversion}

N-version programming \citep{avizienis1985nversion} proposed multiple
independently-developed implementations as a path to software fault
tolerance: if $N$ implementations are independent in their failure modes,
voting among them can mask single-implementation faults. The approach was
largely abandoned in practice because the cost of developing $N$
independent implementations of any substantial system was prohibitive,
and because empirical work \citep{knight1986independence} suggested that
independently-developed implementations often shared correlated failure
modes anyway.

AI-paired engineering changes the cost argument fundamentally. With a
shared specification and AI handling most of the per-port implementation
work, $N$ implementations across $N$ different languages and frameworks
become feasible for a single developer. The independence concern is
genuinely weakened: the implementations share a specification
(deliberately) and share priors from AI training data (less deliberately).
However, the implementations differ substantially in their use of native
platform features --- UI frameworks, rendering models, state management
idioms --- which provides a degree of platform-induced diversity that
human-team N-version programs of the 1980s could not easily achieve,
since those programs typically used the same language and platform across
the $N$ versions.

This is not a claim that N-version differential testing is now formally
equivalent to N-version programming as Avizienis proposed it. It is a
claim that the \emph{economic argument} against N-version programming has
shifted, and that a methodology that uses $N$ implementations as a
differential-testing layer rather than a fault-tolerance voting layer is
now practical for a single developer.

\section{AI-paired engineering in practice}
\label{sec:ai}

\subsection{The dialog-and-review loop}

The methodology centers on two iterative loops. The outer loop turns prose
design documents into the YAML specification: a design doc describes the
objective in English, an analysis prompt surfaces inconsistencies and
missing detail, conversation clarifies the design, and the design doc is
rewritten with the clarifications. Only then does YAML get written. The
analysis prompt used throughout the project:

\begin{quote}
\emph{Please read and understand these requirements. Analyze them for
inconsistencies and completeness. Make suggestions for improvements. Rank
your responses in priority from high to low, and giving each a number.
What are the benefits? What are the downsides? Be ready for a deep dive
into any of the suggestions.}
\end{quote}

The inner loop refines the implementation: a feature is implemented in
YAML, ports are updated, manual tests reveal divergences, and the spec or
implementations are refined. Periodically --- typically every one to two
weeks --- a codebase-wide review is invoked:

\begin{quote}
\emph{Review the entire codebase and evaluate it for clarity,
maintainability, efficiency, complexity, safety, test coverage, pattern
consistency, conformity with style conventions, functional equivalence
across languages, and anything else of importance. Make suggestions for
improvements, ranking them in priority from high to low, and giving each
a number.}
\end{quote}

Both prompts produce ranked lists. The developer picks items to deep-dive.
The ranked-numbered-deep-dive structure externalizes prioritization,
prevents the conversation from drifting into the first item the AI
surfaces, and produces a record that can be revisited.

\begin{figure}[h]
\centering
\begin{tikzpicture}[
  node distance=0.45cm,
  proc/.style={rectangle, draw, rounded corners,
               minimum width=5cm, minimum height=0.65cm,
               font=\small, align=center, inner sep=2pt},
  arrow/.style={->, thick, >=Stealth},
  feedback/.style={->, thick, >=Stealth, gray!70!black, dashed}
]
\node[proc] (n1) {Design doc (prose)};
\node[proc, below=of n1] (n2) {Analysis prompt};
\node[proc, below=of n2] (n3) {Clarifying conversation};
\node[proc, below=of n3] (n4) {Revised design doc};
\node[proc, below=of n4] (n5) {YAML specification};
\node[proc, below=of n5] (n6) {Implement in $N$ ports};
\node[proc, below=of n6] (n7) {Automated + manual tests};
\node[proc, below=of n7] (n8) {Codebase review (every 1--2 weeks)};

\draw[arrow] (n1) -- (n2);
\draw[arrow] (n2) -- (n3);
\draw[arrow] (n3) -- (n4);
\draw[arrow] (n4) -- (n5);
\draw[arrow] (n5) -- (n6);
\draw[arrow] (n6) -- (n7);
\draw[arrow] (n7) -- (n8);

% Refine loop: tests back to spec
\draw[feedback]
  (n7.east) -- ++(1.2,0)
  node[midway, above, font=\scriptsize, black]{refine}
  |- (n5.east);

% Review feedback to design doc
\draw[feedback]
  (n8.west) -- ++(-1.2,0)
  |- (n1.west);

% Cross-cutting note
\node[below=0.6cm of n8, text width=7.5cm, align=center,
      font=\footnotesize\itshape]
  {Cross-cutting: file-based memory (57 entries),
   \texttt{CLAUDE.md} project rules};
\end{tikzpicture}
\caption{Methodology workflow. The outer loop turns prose design into
YAML specification; the inner loop refines the specification through
manual testing across ports. Memory and the periodic codebase-review
prompt cross-cut both loops.}
\label{fig:workflow}
\end{figure}
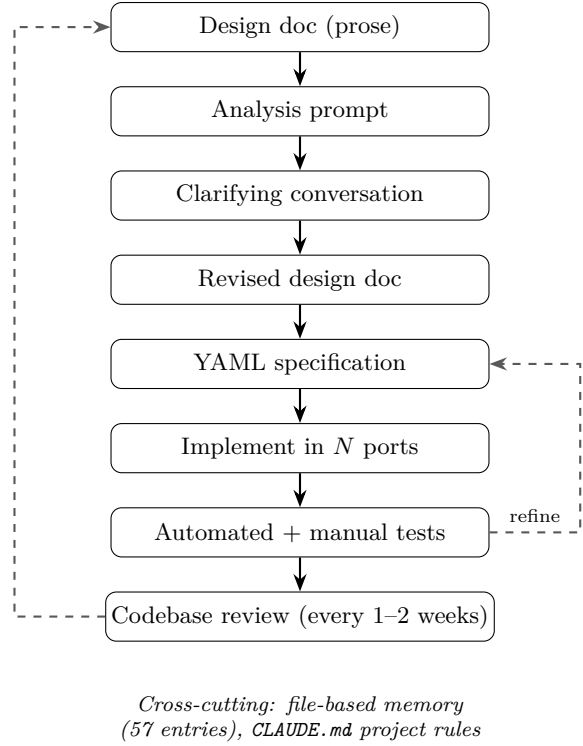

\subsection{Memory as persistent state}

Claude Code provides a file-based memory system that persists across
sessions. Across the seven weeks of the project, 57 memory files
accumulated, falling into four categories: \emph{user} (the developer's
background and preferences), \emph{feedback} (corrections the AI should
not repeat), \emph{project} (state of in-progress features and decisions),
and \emph{reference} (where to find external information). Memory entries
are written by the AI when explicit feedback is given and read at the
start of new sessions.

Memory is not optional. Without it, every session begins fresh, and
decisions accumulated over weeks must be re-established. With it, the AI
carries forward intent --- features deferred, patterns preferred, bugs
seen and avoided. Memory is the closest analog in AI-paired development
to the accumulated context that experienced colleagues bring to a
project.

\begin{figure}[h]
\centering
\begin{lstlisting}
---
name: Verify Swift @ObservedObject claims against ownership chain
description: Before acting on a subagent's flag of @ObservedObject vs
  @StateObject, trace the ownership chain to the construction site
type: feedback
---
Subagent code reviews of SwiftUI frequently misflag `@ObservedObject` as
"should be `@StateObject`". Don't propagate that claim without checking
who constructs the object.

**Why:** A subagent flagged ContentView.swift:110 as wrong, but
JasApp.swift:53 owns the object via `@StateObject` and passes it down -
the textbook pattern. Swapping ContentView to `@StateObject` would have
been an actual regression.

**How to apply:** When a review flags `@ObservedObject`, find the object's
construction site. If the owner uses `@StateObject` and passes the value
down, the receiver's `@ObservedObject` is correct.
\end{lstlisting}
\caption{Example memory entry
(\texttt{feedback\_swift\_ownership\_review.md}). The frontmatter
classifies the entry type and gives it a stable identity; the body
states the rule, the precipitating incident, and the verification
procedure. This is the entry quoted in Section~\ref{sec:delegation}.}
\label{fig:memory}
\end{figure}

\subsection{Manual testing as the dominant remaining cost}

Manual testing was the slowest part of the project. Automated tests catch
state-level divergence and regression, but the visual and behavioral
details of UI rendering --- focus order, animation timing, theme color
application, scroll-bar positioning, gesture recognition --- require
human inspection. The project's manual-testing protocol uses transcript
files (e.g., \texttt{transcripts/COLOR\_TESTS.md}) with numbered
scenarios that an operator runs across each port and marks with pass
dates. The transcripts serve as both a test specification and a
historical record of cross-port differences.

Most manual testing sessions were routine: scenarios pass, dates get
logged, the next feature is queued. A smaller fraction surfaced
challenging problems whose cause was non-obvious. These were generally
not fundamental flaws in the methodology; they were details that the
specification had not yet been forced to disambiguate.

\subsection{Delegation patterns}
\label{sec:delegation}

Claude Code supports delegation to subagents --- auxiliary AI invocations
with their own context windows. Subagents helped most when the task was
independent and well-scoped: finding every place where a symbol is
referenced, summarizing the current state of a feature across all ports.
They hurt when the task required cross-file design judgment or
ownership-chain reasoning, where the subagent could not see the
surrounding context that justified an existing pattern. A memory entry
that survived from early in the project captures the recurring failure
mode:

\begin{quote}
\emph{Subagents over-flag Swift \texttt{@ObservedObject} patterns as bugs.
Verify against the ownership chain before acting; subagents frequently
misflag correct patterns.}
\end{quote}

The general rule: subagent summaries describe intent, not effect. Verify
the diff that resulted from a subagent's recommendation, not the
summary.

\subsection{Honest failure modes}

Several AI failure modes recurred throughout the project, even with the
safeguards described in Sections~\ref{sec:spec} and~\ref{sec:nimpl} in
place.

\emph{Long-context drift.} In sessions extending beyond a few hours, the
AI gradually loses track of earlier decisions, re-introducing patterns
that were rejected earlier or suggesting ``improvements'' to code that was
deliberately load-bearing. Mitigated by memory entries for durable
decisions and periodic session restarts.

\emph{Confident hallucinations of symbols and file paths.} The AI
sometimes invents file names, functions, or APIs that do not exist,
presented with high confidence. The mitigation is habitual cross-checking
via \texttt{grep} before acting on AI-supplied references.

\emph{Optimistic completion summaries.} The AI occasionally reports a task
as complete when the implementation is partial or has introduced
regressions. The mitigation is direct diff inspection rather than
trusting the summary.

\emph{Spec underspecification surfacing late.} When the YAML specification
permits multiple reasonable interpretations, the AI picks one confidently
without flagging the ambiguity. This is not a flaw of the AI per se; it
is a property that means underspecification surfaces only at integration
time.

\emph{\texttt{--dangerously-skip-permissions} trade-off.} The project was
developed primarily with this flag enabled, meaning the AI could modify
files and run commands without per-action confirmation. This was
essential for development velocity. The safety net was the combination of
automated tests, manual testing, and $N$ implementations: a destructive
change would either break a test or produce a visible divergence between
ports. The trade-off is operational; it is not appropriate for every
project.

\subsection{Tooling implications and recommendations}

The features that mattered most for this project, in approximate order of
impact: persistent file-based memory across sessions; project-level
instructions in \texttt{CLAUDE.md}; file-editing tools that the AI can
use directly; the \texttt{--dangerously-skip-permissions} trust mode; a
long context window (1M tokens) that allows sessions to proceed without
re-explaining state; and the slash-command and task-tracking facilities
of Claude Code.

The features I would most want from future iterations: stronger
guardrails against hallucinated symbols, perhaps via tighter integration
with language servers; better cross-session retention without explicit
memory writes; collaborative shared memory across multiple developers;
and more principled handling of ``design fork'' decision points, where
the AI is currently inclined to choose confidently rather than surface
the choice.

These observations are tuned to Claude Opus 4.6 and 4.7 and to Claude
Code as it existed in April--May 2026; Section~\ref{sec:limits}
discusses this snapshot limitation.

\section{Evidence and results}
\label{sec:evidence}

\subsection{Project artifacts}

The artifact comprises five working ports of a vector illustration
application plus a shared specification. Total source line counts at the
time of writing:

\begin{table}[h]
\centering
\caption{Lines of code by component.}
\begin{tabular}{llr}
\toprule
Component & Language & LOC \\
\midrule
jas\_dioxus & Rust & 95{,}371 \\
JasSwift & Swift & 71{,}883 \\
jas\_ocaml & OCaml & 69{,}043 \\
jas & Python (PySide6) & 59{,}069 \\
jas\_flask & Python + HTML & 5{,}239 \\
Shared specification & YAML & 22{,}866 \\
Shared interpreter & Python & 12{,}569 \\
\midrule
\textbf{Total} & & \textbf{$\sim$336{,}000} \\
\bottomrule
\end{tabular}
\end{table}

The five ports together contain approximately 4{,}600 automated test
functions and 36 manual-test transcript files. The Color Panel alone
defines 98 numbered manual scenarios. The application implements 27
tools, 14 panels, and 22 dialogs, declared in the YAML specification and
rendered through the per-port renderers described in
Section~\ref{sec:spec}.

The repository history records 1{,}807 commits over 48 calendar days,
with 40 of those days containing at least one commit (an activity rate
of 83\% across the calendar span). Cross-port divergence is explicitly
noted in 32 commit messages and is implicit in many more --- propagation
merges following the standard order (Flask $\rightarrow$ Rust
$\rightarrow$ Swift $\rightarrow$ OCaml $\rightarrow$ Python) carry the
divergences observed in earlier ports forward into later ones.

\subsection{Effort}

The author worked on the project during evening hours outside
professional employment. The 48-day calendar span comprises roughly
seven weeks, with active days clustering on weeknights and weekends.
The total developer effort is difficult to measure precisely;
commit-time clustering and the calendar pattern of active days suggest
an estimate of 120 to 160 evening hours. This is consistent with the
author's recollection of roughly three to four hours per active evening
across the seven weeks.

\subsection{Comparison to conventional development}

Mature vector illustration applications of comparable scope represent
decades of team development. Adobe Illustrator has shipped continuously
since 1987; Inkscape has been actively developed since 2003. Affinity
Designer, the closest commercial single-team comparison, has been in
development since the early 2010s and continues to be produced by a
sustained team. None of these is directly commensurable with a
single-developer, single-language port; the comparison is approximate.

A more constrained comparison is to estimate what one developer working
full-time without AI assistance would require to produce a single port
at the scope reported here --- paths, text, layers, transforms, document
model, panels, dialogs, tools, undo, save and restore, PDF export.
Conservative estimates for a single working port of comparable scope, by
a competent developer using conventional tools and frameworks, run from
one to two developer-years. Five such ports, even sharing a single
specification, would conventionally represent multiple developer-years.

The artifact described in this paper represents approximately 120 to 160
developer-hours, distributed across five ports over seven calendar
weeks. This is not a controlled comparison and the productivity ratio is
not a measurement. It is, however, consistent with a claim that
AI-paired engineering under the conditions of Sections~\ref{sec:spec}
and~\ref{sec:nimpl} changes the scope of work feasible for a single
developer by at least one and possibly two orders of magnitude.

\section{Limitations}
\label{sec:limits}

This is a single-developer case study, and several limitations apply. I
name them explicitly here in the order most likely to affect
interpretation of the central claims.

\paragraph{Single developer, no controlled comparison.} The case study
has $N=1$ developer and no control group, no parallel non-AI baseline,
and no replication by other developers. The 120 to 160 developer-hour
estimate is derived from commit timestamps and the calendar pattern of
active days; it is not tracked time, and the conversion from ``active
hour-bucket'' to ``focused hour'' carries some uncertainty. Comparisons
to mature applications (Section~\ref{sec:evidence}) are scope anchors,
not commensurable baselines.

\paragraph{Scope vs.\ mature applications.} The artifact implements a
substantial subset of the feature set found in mature vector
illustration applications but is not complete relative to them. It lacks
gradient mesh, advanced text shaping at the level of professional
typesetting engines, raster effects, full SVG round-trip fidelity,
plugins, and color management beyond basic profiles. Pixel-level visual
conformance across ports was not formally measured; the implementations
look similar by inspection but exact conformance has not been verified.

\paragraph{Developer expertise as confound.} I have used vector
illustration applications since 1990 and have substantial prior
experience in OCaml and Python; less so in Swift and Rust. Domain
familiarity reduced design uncertainty during specification writing.
Language expertise reduced friction in reviewing and correcting
AI-generated code. A developer without this background, attempting the
same methodology, might find the productivity gain smaller; the
methodology has not been tested under that condition.

\paragraph{Generalization limits.} The methodology depends on a domain
that can be described declaratively. UI panels, dialogs, and tool
behavior fit this constraint; domains with deep imperative interaction
(game engines, real-time graphics pipelines, kernel code) may not. The
N-version revival claim is contingent on AI being capable of producing
per-port mechanical work at low cost; this is contingent on AI capability
at any given moment. Manual testing scales linearly with $N$, and at much
larger $N$ or much larger feature surface this could become the dominant
cost. AI-generated implementations also share priors from training data,
making them less independent than human-team N-version programs.

\paragraph{AI capability is a moving target.} Results reported here used
Claude Opus 4.6 and 4.7 as accessed via Claude Code during April and May
2026. Workflow patterns described --- the dialog loop, the
codebase-review prompt, persistent memory,
\texttt{--dangerously-skip-permissions} --- are tuned to current model
capability and to current Claude Code tooling. Both are changing rapidly.
A less-capable model or a less-supportive client would likely introduce
friction the present methodology does not exhibit. The case study is a
snapshot; replication six to twelve months later may yield different
results.

\section{Related work}
\label{sec:related}

\paragraph{Executable specifications.} Spec-as-contract has a long
lineage in language semantics and compilers --- POSIX conformance,
ECMA-262 / test262, the WebAssembly reference interpreter
\citep{haas2017webassembly}, and the K-Framework
\citep{rosu2010koverview}. These projects use an executable spec to
anchor multiple implementations, typically with one ``reference''
implementation and a conformance test suite. This work shares the
spec-as-contract pattern but applies it to application-level semantics
--- UI, reactive state, dialog flows --- and shifts from
one-reference-plus-tests to $N$ peer implementations developed in
parallel.

\paragraph{Differential testing and N-version programming.} Differential
testing \citep{mckeeman1998differential} has succeeded in compiler
validation \citep{yang2011csmith}, where multiple independent compilers
already exist. N-version programming \citep{avizienis1985nversion}
proposed multiple independently-developed implementations for software
fault tolerance, but was largely abandoned because the cost of
developing $N$ independent implementations was prohibitive; empirical
work \citep{knight1986independence} further showed that
independently-developed implementations often shared correlated failure
modes. \textbf{AI fundamentally changes that economic argument.} This
paper is, to my knowledge, the first existence proof that $N$
implementations developed against a shared specification, with AI
handling most of the per-port mechanical work, are feasible for a single
developer. The $N$ implementations here span five distinct programming
languages and UI frameworks, providing implementation diversity that may
exceed what was achievable in single-language N-version projects of the
1980s.

\paragraph{AI productivity literature.} Controlled studies of
AI-assisted development \citep{peng2023copilot,ziegler2024copilot}
measure narrow tasks and report 0--80\% productivity gains. A recent
rigorous study \citep{metr2025productivity} found negative effects on
experienced developers working on familiar code, complicating the
picture. This paper complements those studies with a project-scale case
report measuring not per-task speed but feasibility of scope. The
methodologies measure different things; both contribute to understanding
when and how AI is useful.

\paragraph{Comparable applications.} Vector illustration applications
represent decades of development by sustained teams. Adobe Illustrator
\citep{pfiffner2003adobe} has been developed since 1987; Inkscape
\citep{inkscape}, the leading open-source comparison, since 2003.
Commercial alternatives such as Affinity Designer and Figma also
represent multi-developer-year efforts by sustained teams. The artifact
described here does not match these in feature completeness; the
contribution is to demonstrate that a substantial subset is achievable
in approximately 120 developer-hours across five platforms by a single
developer.

\section{Conclusion}
\label{sec:conclusion}

This paper reports a single-developer case study in AI-paired software
engineering: five working ports of a vector illustration application,
sharing one executable YAML specification, built across approximately
120 evening hours over seven calendar weeks. The methodology relies on
two safeguards --- a precise executable specification that consolidates
design decisions, and parallel implementations that function as a
built-in differential-testing layer. I argue these safeguards together
make feasible for a single developer a scope of work that would
conventionally require multiple developer-years.

The methodology generalizes to other domains that can be described
declaratively. It is less clear how far it extends into domains with
deep imperative interaction or into projects without a clear notion of
platform-spanning equivalence. The economic argument for $N$
implementations --- once impractical, now feasible --- appears to me the
most durable claim of this work; the specific productivity numbers will
shift as AI capability shifts.

For a reader interested in replicating the approach, the minimal recipe
is: choose a domain that can be described declaratively; pick $N$ ports
across distinct languages and UI frameworks ($N=3$ captures most of the
correctness benefit); use AI for per-port mechanical work; iterate
between a prose design document and the executable specification via the
analysis prompt (Section~\ref{sec:ai}); cross-test ports manually using
numbered transcript files; and maintain memory across sessions for
accumulated decisions. The artifact and accompanying methodology
documents are available at \url{https://github.com/jyh/jas}.

\bibliographystyle{plainnat}
\bibliography{references}

\end{document}